\documentclass[apj]{emulateapj}

\usepackage{natbib}
\usepackage{amsmath,amssymb}
\usepackage{mathptmx}

\usepackage{graphicx}

\newcommand{\todo}{\ifmmode {\Huge \bullet} \else {\Huge$\bullet$}\fi}
\newcommand{\til}{\ifmmode \sim \else $\sim$\fi}

\newcommand{\kms}	{\ifmmode {\rm km\,s}^{-1} \else km\,s$^{-1}$\fi}
\newcommand{\ergs}	{\ifmmode {\rm erg\,s}^{-1} \else erg s$^{-1}$\fi}
\newcommand{\Msun}{\ifmmode M_{\odot} \else $M_{\odot}$\fi}
\newcommand{\mpyr}{\ifmmode \Msun\,{\rm yr}^{-1} \else $\Msun\,{\rm yr}^{-1}$\fi}

\newcommand{  \Halpha   }{\ifmmode {\rm H}\alpha \else H$\alpha$\fi}
\newcommand{  \ha   	}{\ifmmode {\rm H}\alpha \else H$\alpha$\fi}
\newcommand{  \Hbeta    }{\ifmmode {\rm H}\beta \else H$\beta$\fi}
\newcommand{  \hb    	}{\ifmmode {\rm H}\beta \else H$\beta$\fi}
\newcommand{  \CIV      }{\ifmmode {\rm C}\,\textsc{iv}\,\lambda1549 \else C\,\textsc{iv}\,$\lambda1549$\fi}

\newcommand{ \fwhb  }{\ifmmode {\rm FWHM}\left(\hb\right) \else FWHM(\hb)\fi}
\newcommand{  \lamLlam  }{\ifmmode \lambda L_{\lambda} \else $\lambda L_{\lambda}$\fi}
\newcommand{  \Lop      }{\ifmmode L_{5100} \else $L_{5100}$\fi}
\newcommand{\fbol}{\ifmmode f_{\rm bol} \else $f_{\rm bol}$\fi}
\newcommand{\fbolwv}{\ifmmode f_{\rm bol}\left(\lambda\right) \else $f_{\rm bol}\left(\lambda\right)$\fi}
\newcommand{\fbolopt}{\ifmmode f_{\rm bol}\left(5100{\rm \AA}\right) \else $f_{\rm bol}\left(5100{\rm \AA}\right)$\fi}

\newcommand{  \mbh      }{\ifmmode M_{\rm BH} \else $M_{\rm BH}$\fi}
\newcommand{  \lledd    }{\ifmmode L/L_{\rm Edd} \else $L/L_{\rm Edd}$\fi}
\newcommand{  \Lbol     }{\ifmmode L_{\rm bol} \else $L_{\rm bol}$\fi}

\newcommand{  \tgrow     }{\ifmmode t_{\rm growth} \else $t_{\rm growth}$\fi}
\newcommand{  \tUni      }{\ifmmode t_{\rm U} \else $t_{\rm U}$\fi}
\newcommand{  \Mbhdot	}{\ifmmode \dot{M}_{\rm BH} \else $\dot{M}_{\rm BH}$\fi}
\newcommand{  \Maddot	}{\ifmmode \dot{M}_{\rm AD} \else $\dot{M}_{\rm AD}$\fi}

\newcommand{  \as	}{\ifmmode a_{\rm *} 		\else $a_{\rm *}$\fi}
\newcommand{  \avec	}{\ifmmode \vec{a}_{\rm *} 	\else $\vec{a}_{\rm *}$\fi}
\newcommand{  \re	}{\ifmmode \eta      	\else $\eta$\fi}
\newcommand{  \mseed    }{\ifmmode M_{\rm seed} \else $M_{\rm seed}$\fi}

\newcommand{  \NhiM   }{48}
\newcommand{  \MhiM   }{\ifmmode 3\times10^{9} \else $3\times10^{9}$	\fi}
\newcommand{  \NexM   }{17}
\newcommand{  \MexM   }{\ifmmode 10^{10} 	\else $10^{10}$		\fi}

\shorttitle{High Spins for the Most Massive Black-Holes at $z\sim1.5-3.5$}
\shortauthors{Trakhtenbrot}

\begin{document}

\title{The Most Massive Active Black-Holes at $z\sim1.5-3.5$ Have High Spins and Radiative Efficiencies}
\slugcomment{Published 13 June 2014, as ApJ, 789, L9}

\author{Benny Trakhtenbrot\altaffilmark{1}}
\affiliation{Institute for Astronomy, ETH Zurich, 
Wolfgang-Pauli-Strasse 27, 
CH-8093 Zurich, Switzerland; \href{mailto:benny.trakhtenbrot@phys.ethz.ch}{benny.trakhtenbrot@phys.ethz.ch}}
\altaffiltext{1}{Zwicky Fellow}


\begin{abstract}
The radiative efficiencies ($\eta$) of 72 luminous unobscured active galactic nuclei at $z\sim1.5-3.5$, powered by some of the most massive black holes (BHs), are constrained.
The analysis is based on accretion disk (AD) models, which link the continuum luminosity at rest-frame optical wavelengths and the BH mass (\mbh) to the accretion rate through the AD, \Maddot.   
The data are gathered from several literature samples with detailed measurements of the \hb\ emission line complex, observed at near-infrared bands.
When coupled with standard estimates of bolometric luminosities (\Lbol), the analysis suggests high radiative efficiencies, with most of the sources showing $\eta>0.2$ - that is, higher than the commonly assumed value of 0.1, and the expected value for non-spinning BHs ($\eta=0.057$).
Even under more conservative assumptions regarding \Lbol\ (i.e., $\Lbol=3\times\Lop$), most of the extremely massive BHs in the sample (i.e., $\mbh\gtrsim3\times10^9\,\Msun$) show radiative efficiencies which correspond to very high BH spins (\as), with typical values well above $\as\simeq0.7$.
These results stand in contrast to the predictions of a ``spin-down'' scenario, in which a series of randomly oriented accretion episodes lead to $\as \sim 0$.
Instead, the analysis presented here strongly supports a ``spin-up'' scenario, which is driven by either prolonged accretion or a series of anisotropically oriented accretion episodes.
Considering the fact that these extreme BHs require long-duration or continuous accretion to account for their high masses, 
it is argued that the most probable scenario for the super-massive BHs holes under study is that of an almost continuous sequence of randomly yet not isotropically oriented accretion episodes.
\end{abstract}

\section{Introduction}
\label{sec:intro}

The spins of super-massive black holes (SMBHs) play a major role in their growth, and hold hints to their evolution.
According to standard accretion disk (AD) theory of active galactic nuclei (AGNs), the spin of the SMBH (in dimensionless units, \as) determines the radius of innermost stable orbit within the AD, and thus the radiative efficiency of the accretion process, $\eta\equiv \Lbol/\Maddot c^2$, where \Maddot\ is the accretion rate through the AD and \Lbol\ is the emergent bolometric luminosity. 
Thus, the spin determines the SMBH mass growth efficiency.

Several numerical studies investigated the evolution of \as, assuming that the SMBH grows through a series of coalescence and/or accretion events.
It was shown that if the different accretion episodes have random and isotropic orientations with regard to the SMBH, it is expected to ``spin down'', reaching $\as\simeq0$ \cite[e.g.,][D13 hereafter]{Volonteri2007_spin,Berti2008,King2008_spindown,Dotti2013}.
The study of \cite{King2008_spindown} showed that a SMBH can reach $\as\lesssim0.2$, while growing to a mass of $\mbh\sim10^{10}\,\Msun$, within $\sim$600 Myr.
Alternatively, if the SMBH grows mainly through one prolonged accretion episode, it is expected to ``spin-up'' and reach $\as\simeq1$.
The recent study of D13 showed that even a simple anisotropy in the orientations of accretion episodes may result in high spins for the most massive black holes (BHs).
The most significant differences between the ``spin-down'' and ``spin-up'' scenarios relate to the spins of the most massive BHs, with $\mbh\gtrsim10^9\,\Msun$, as these experience more accretion and coalescence events. 
To test the relevance of these two spin evolution scenarios, the spins of extremely massive BHs have to be estimated, preferably at different redshifts.

Measuring BH spins is, however, notoriously challenging. 
One such method is based on observing the relativistically broadened K$\alpha$ line, which probes the innermost regions of the AD (see \citealt{Reynolds2013_Ka_spins} for a recent review).
This method was applied in $\sim$20 low-redshift AGNs, with $\mbh\sim10^6-3\times10^8\,\Msun$, providing measurements that cover the entire range of $-0.99 \lesssim \as \lesssim0.99$, 
with large uncertainties (e.g., \citealt{Brenneman2011} and \citealt{Patrick2012}).
The K$\alpha$ method can hardly be applied to more massive BHs, which are mostly inactive in the local universe \cite[e.g.,][TN12 hereafter]{Schulze2010_BHMF,Shen_Kelly2012,TrakhtNetzer2012_Mg2}.
Indirect and statistical evidence for BH spin was also deduced from the cumulative evolution of the AGN population \cite[the so-called Soltan argument; see][]{Soltan1982}.
This approach yields $\left<\eta\right>\simeq0.1$, corresponding to $\as\simeq0.67$ \cite[e.g.,][and references therein]{Marconi2004}, with some evidence for higher spins at higher redshifts \cite[e.g.,][]{Wang2009_rand_acc,Shankar2010_eff,Li2012}.  
Obviously, this approach cannot probe the spins of \emph{individual} objects.

This Letter uses insights from AD theory and basic observables of unobscured AGNs, to estimate the radiative efficiencies and corresponding BH spins of a large sample of luminous, high redshift AGNs, which were shown to be powered by some of the most massive BHs in the universe ($\mbh\gtrsim3\times10^9\,\Msun$). 
This work assumes $\Omega_{\Lambda}=0.7$, $\Omega_{\rm M}=0.3$, and $H_{0}=70\,\kms\,{\rm Mpc}^{-1}$.

\begin{figure*}[ht!]
\includegraphics[width=0.48\textwidth]{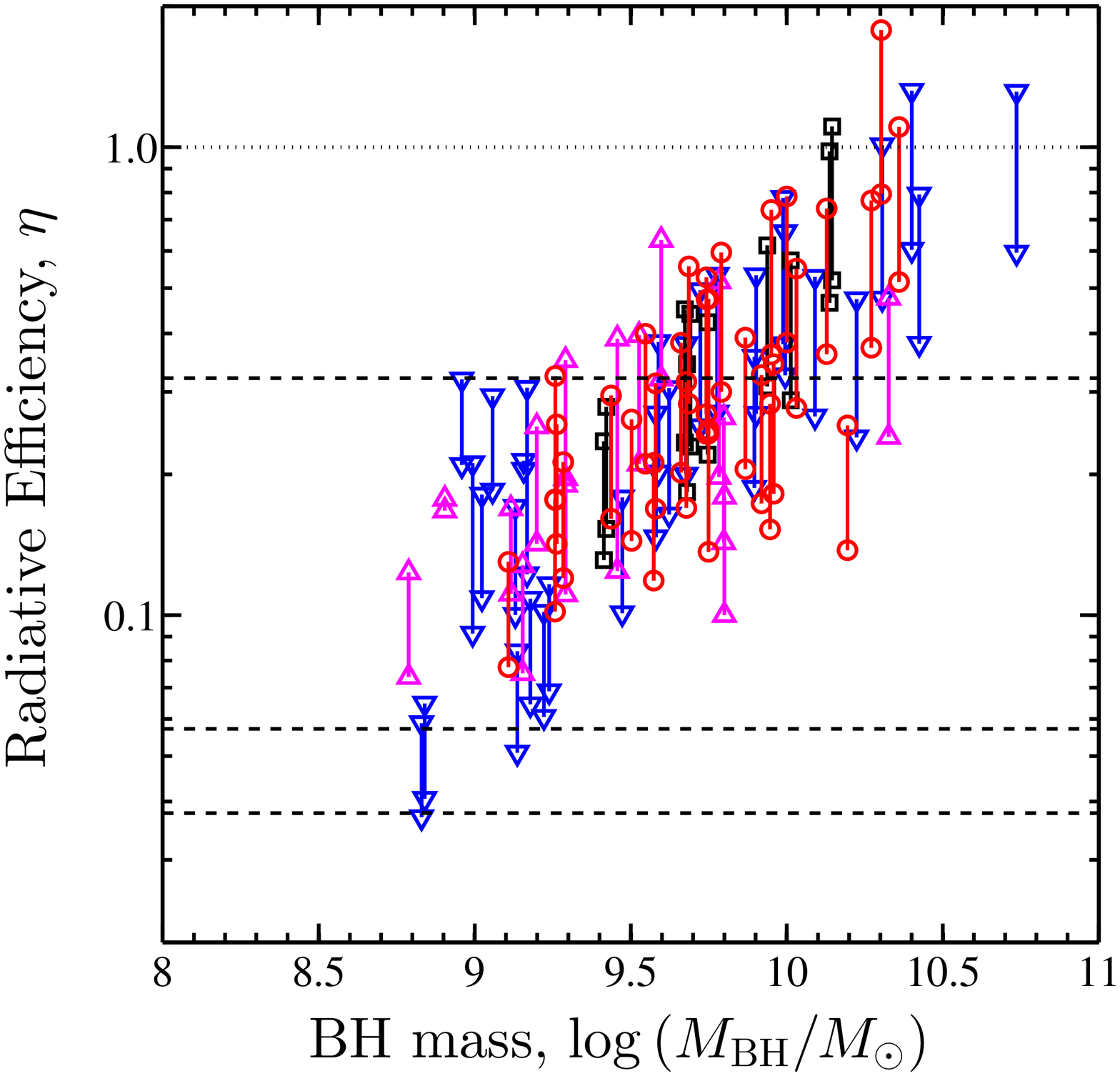} 
\hspace{0.3cm}
\includegraphics[width=0.48\textwidth]{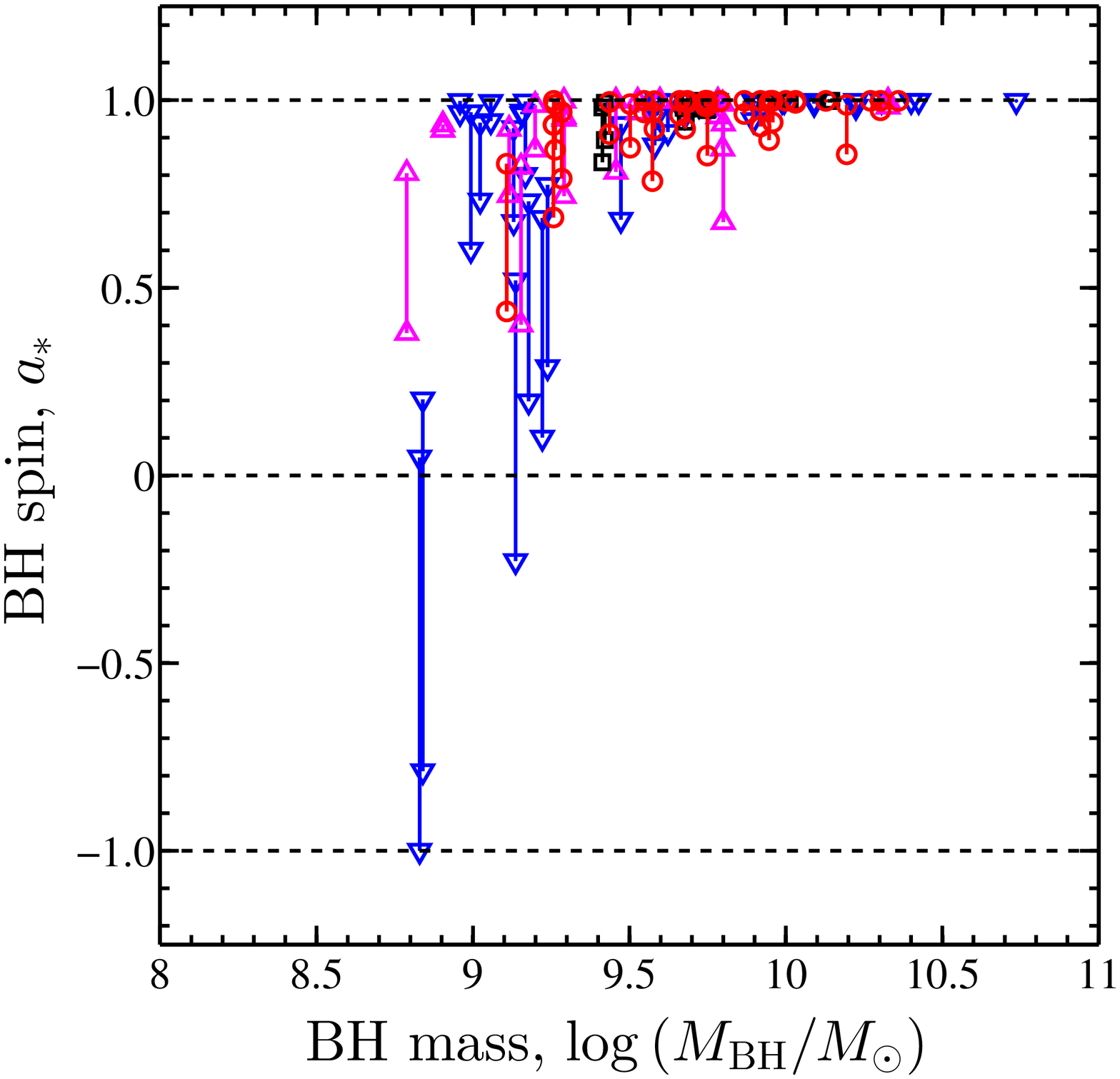} 
\caption{
Deduced radiative efficiencies ($\eta$; \emph{left}), and BH spins (\as; \emph{right}),  vs. \mbh, for the high-redshift AGNs under study.
In both panels, the different symbols correspond to the samples used: the combined sample of \cite{Shemmer2004} and \citet[][blue and magenta triangles for sources at $z\simeq2.4$ and $\simeq3.3$, respectively]{Netzer2007_MBH}, \citet[][red circles]{Marziani2009} and \citet[][black squares]{Dietrich2009_Hb_z2}. 
For each source, two values are plotted, based on Equation~(\ref{eq:mdot_ad}) (with \Lop) and on the prescription of \citet[][their Equation (9)]{DavisLaor2011_AD}.
Dashed horizontal lines mark specific spins and corresponding radiative efficiencies, of $\as=-1$, 0, and 0.998. 
The dotted line in the left panel marks the (physically unrealistic) case $\eta=1$.
}
\label{fig:eta_a_vs_mbh_both_methods}
\end{figure*}

\section{Samples and Analysis}
\label{sec:samples_analysis}

The goal of the present study is to derive constraints on the radiative efficiencies of the ADs powering a large sample of luminous, unobscured AGNs at $1.5\lesssim z \lesssim3.5$.
This is done by estimating the mass accretion rate (\Maddot) and emergent bolometric luminosities (\Lbol), which are then combined to yield the radiative efficiency, $\eta \equiv \Lbol/\Maddot\,c^2$.
This approach follows the one presented in \cite{BianZhao2003_spin}, who focused on low-redshift, low-\mbh\ sources, 
and further used for large samples of $z<2$ AGNs \cite[e.g.,][NT14 hereafter]{Wu2013_eff,NetzerTrakht2014_slim}

The samples and data used here are presented in detail in the studies by
\cite{Shemmer2004}, \cite{Netzer2007_MBH}, \cite{Dietrich2009_Hb_z2} and \cite{Marziani2009}.
In order to focus on high-redshift SMBHs, this study analyzes 72 sources at $z>1.3$, omitting a total of thirteen sources from the \cite{Marziani2009} and \cite{Dietrich2009_Hb_z2} samples. 
This choice does not affect the results presented in this work.
These data sets include near-infrared (IR) spectra that cover the \hb\ broad emission line and the continuum emission at rest-frame wavelength of 5100 \AA\ (\Lop), which can be used to provide reliable estimates of \mbh, using ``virial'' \mbh\ estimators \cite[see][for a review]{Shen2013_rev}. 
\Lop\ can also be used to estimate \Lbol\ (see below).
The spectra of the different samples were analyzed in a very similar way, generally following the prescriptions presented in TN12, providing consistent measurements of \Lop\ and \mbh.
An important feature of the sample studied here is that it includes some of the most massive BHs known to date, with $\mbh>\MhiM\,\Msun$ for \NhiM\ (64\%) of the sources, and $\mbh>\MexM\,\Msun$ for \NexM\ (24\%).

The accretion rates through the ADs, \Maddot, were estimated in several ways, all based on the AGN continuum luminosity at the (rest-frame) optical regime. 
The basic assumption is that the long-wavelength power-law tail of the UV-optical spectral energy distribution (SED) of AGNs, predicted by a simple geometrically thin AD model \cite[e.g.,][SS73 hereafter]{ShakuraSunyaev1973}, probes the outer parts of the AD and is mostly unaffected by the BH spin. 
The analytical derivation of \Maddot\ prescriptions, assuming a SS73-like AD, is well established \cite[e.g.,][]{Bechtold1987,Collin2002}. 
The recent study of \citet[][DL11 hereafter]{DavisLaor2011_AD} demonstrated that more elaborate AD models still reproduce similar prescriptions for  \Maddot, given reliable estimates of (rest-frame) optical luminosities and \mbh.
For SMBHs with $\mbh\lesssim10^{9}\,\Msun$ the power-law tail dominates the SED at $\lambda\gtrsim4500$ \AA. 
For such sources ($\sim$20\% of the sample studied here), \Maddot\ can be estimated directly from \mbh\ and \Lop\ following the general prescription:  
\begin{equation}
 \Maddot \simeq 2.4\, \left(\frac{\lambda L_{\lambda,45}}{\cos i}\right)^{3/2}\,\left(\frac{\lambda}{5100{\rm \AA}}\right)^2 \, M_{8}^{-1} \,\,\, \mpyr \,\, ,
 \label{eq:mdot_ad}
\end{equation} 
where $\lambda$ is the wavelength at which the continuum emission is measured, 
$\lambda L_{\lambda,45}\equiv\lamLlam/10^{45}\,\ergs$, 
$M_{8}\equiv\mbh/10^{8}\,\Msun$, 
and $\cos i$ represents the inclination of the AD with regard to the line of sight, assumed here to be $\cos i=0.8$.
The study of DL11 suggests a smaller scaling factor\footnotemark[2] of $\sim1.7$, and shows that this choice may result in systematic offsets of up to $\sim30\%-40\%$ at the high-\mbh\ regime.
\footnotetext[2]{That is, instead of 2.4 in Equation~(\ref{eq:mdot_ad}). See footnote~1 in \cite{LaorDavis2011_WLQs}}
The particular scaling factor in Equation~(\ref{eq:mdot_ad}) was chosen to provide consistent results even for relatively high-\mbh\ sources (see below). 
Due to the need to assume a certain AD model, such uncertainties may underlie any attempt to estimate \Maddot\ for individual AGNs. 
In any case, for the purposes of this work, the expression in Equation~(\ref{eq:mdot_ad}) serves as a conservative estimate for \Maddot\ for the low-\mbh\ sources, as choosing a 
smaller scaling factor would result in even larger values for $\eta$ (and thus for \as).
For consistency with previous work, \Maddot\ was also estimated by the empirically-derived relation of DL11 (their Equation~(9)), based on \Lop, \fwhb, and the TLUSTY AD model.
These estimates provide somewhat lower accretion rates, although the differences are small (see DL11 for more details).
For more massive BHs, with $\mbh\gtrsim3\times10^{9}\,\Msun$, which are of central interest to the present analysis, the power-law tail of the SED is shifted to yet longer wavelengths \cite[e.g., Fig. 1 in][]{LaorDavis2011_WLQs}. 
Therefore, estimates of \Maddot\ from \Lop\ do have an increased dependence on BH spin, as the corresponding flux is emitted from regions of the AD which are closer to the BH.
A comparison of the values given by Equation~(\ref{eq:mdot_ad}) and the results of a grid of AD models for $\mbh=10^{10}\,\Msun$, based on the study of \citet[][provided by H. Netzer, private communication]{Slone2012} indicates that Equation~(\ref{eq:mdot_ad}) may, at the most extreme cases, \emph{under-estimate} \Maddot\ by $\sim$30\% (0.15 dex).
In the context of the present analysis, such extreme cases would result in a corresponding \emph{over-estimation} of \re.
However, these uncertainties are negligible in comparison to the systematic uncertainties on the \mbh\ estimates used here \cite[$\sim$0.3 dex. See, e.g.,][]{Shen2013_rev}.
Considering these large systematic uncertainties, the formal uncertainty on any estimate of \Maddot\ (and thus of \re) may reach $\sim$0.5 dex.

Due to the uncertainties related to high-\mbh\ systems, another set of \Maddot\ estimates was derived by using monochromatic luminosities at longer wavelengths, available from IR surveys, for a total of 58 sources.
In particular, the public Two Micron All Sky Survey (2MASS) and \emph{Wide-Field Infrared Surveys Explorer} (\emph{WISE}) surveys (\citealt{Skrutskie2006_2mass} and \citealt{Wright2010_WISE}, respectively) probe rest-frame wavelengths of $6500<\lambda_{\rm rest}<9500$ \AA.
The emission in this wavelength regime remains only marginally affected by BH spin, even for extremely massive BHs. 
The catalogs of these surveys were queried for counterparts within 3{\hbox{$^{\prime\prime}$}} of the (optical) coordinates of the sources.
Estimates of \Maddot\ were derived only from reliable flux measurements (S/N$>$3) which probe the aforementioned wavelength range.
These ``IR-based'' estimates of \Maddot\ are systematically higher than those obtained by using \Lop\ (by up to a factor of $\sim2$). 
Some of this effect may be due to contamination by the broad \Halpha\ line.

The bolometric luminosities of the sources, \Lbol, were estimated from \Lop, by applying two different types of bolometric corrections, $\fbolopt \equiv\Lbol/\Lop$.
The first approach uses ``standard'' bolometric corrections, which are based on composite SEDs of AGNs \cite[][]{Elvis1994,VandenBerk2001,Richards2006_SED}. 
The present work uses the corrections presented in TN12, which are designed to be consistent with those of \cite{Marconi2004}.\footnote[3]{These corrections are based on an SED model that extends beyond a simple AD, and includes X-ray emission.}
For the sample studied here, these prescriptions provide a range in bolometric corrections of $\fbolopt\sim5.3-6.5$.
These ``standard'' bolometric corrections, however, suffer from two major issues, that may limit their use in the present context.
First, any prescription for \fbolopt\ neglects the scatter of this quantity within the AGN population \cite[e.g.,][]{Elvis1994,Richards2006_SED,Jin2012a,Runnoe2012}.
Second, and more importantly, these empirical corrections rely on composite SEDs of several AGNs, for which the BH spin is unknown. 
These unknown spins are expected to determine the far-UV emission of the AGNs, and thus to affect the integrated \Lbol\ values.
The ``standard'' bolometric corrections thus represent the range of BH spins in the samples used to construct the composite SEDs.
Due to these shortcomings, the analysis in this work also relies on more conservative estimates of \Lbol.
The ``conservative'' estimates of \Lbol\ assumed $\Lbol=3\times\Lop$ (i.e., $\fbolopt=3$), a factor of $\sim2$ lower than the ``standard'' corrections.
This value was chosen to provide strong lower limits on \Lbol, according to our current understanding of AGN SEDs. 
For example, almost all the Palomar-Green quasars studied by DL11, including those with very low deduced \re, have $\fbolopt>3$, and the typical value is actually $\fbolopt\simeq8.7$ \cite[see also][]{Jin2012a,Runnoe2012}.
In terms the commonly used UV-optical SED of an unobscured AGN, with $f_{\nu}\propto\nu^{-1/2}$ \cite[][]{VandenBerk2001}, the choice of $\fbolopt=3$ is equivalent to calculating \Lbol\ by integrating only between $\sim$1200 and $\sim$6500 \AA, that is, completely ignoring the ionizing continuum.
Thus, the conservative estimates of \Lbol\ provide strong lower limits on \re\ and \as, as the choice of 
any higher (and more reasonable) \fbolopt\ would result in yet higher values for these quantities.

The various estimates of \Maddot\ and \Lbol\ were combined to estimate radiative efficiencies, as defined above, and translated to \as, following standard prescriptions \cite[e.g.,][]{NovikovThorne1973}.
These translations of \re\ to \as\ are highly non linear. This means that the uncertainties on \Maddot\ (and thus \re), of order $\sim$40\%, may result in much smaller uncertainties on \as.
As an example, for a BH with a deduced spin of $\as=0.85$, an over-estimation of \re\ by 40\% would imply that the true spin is $\as=0.65$, i.e., still a fast spinning BH.
The systematic uncertainties on \Maddot\ estimates remain, however, the largest caveat of the present analysis, along with the large systematics on \mbh\ estimates.

\begin{figure*}
\includegraphics[width=0.48\textwidth]{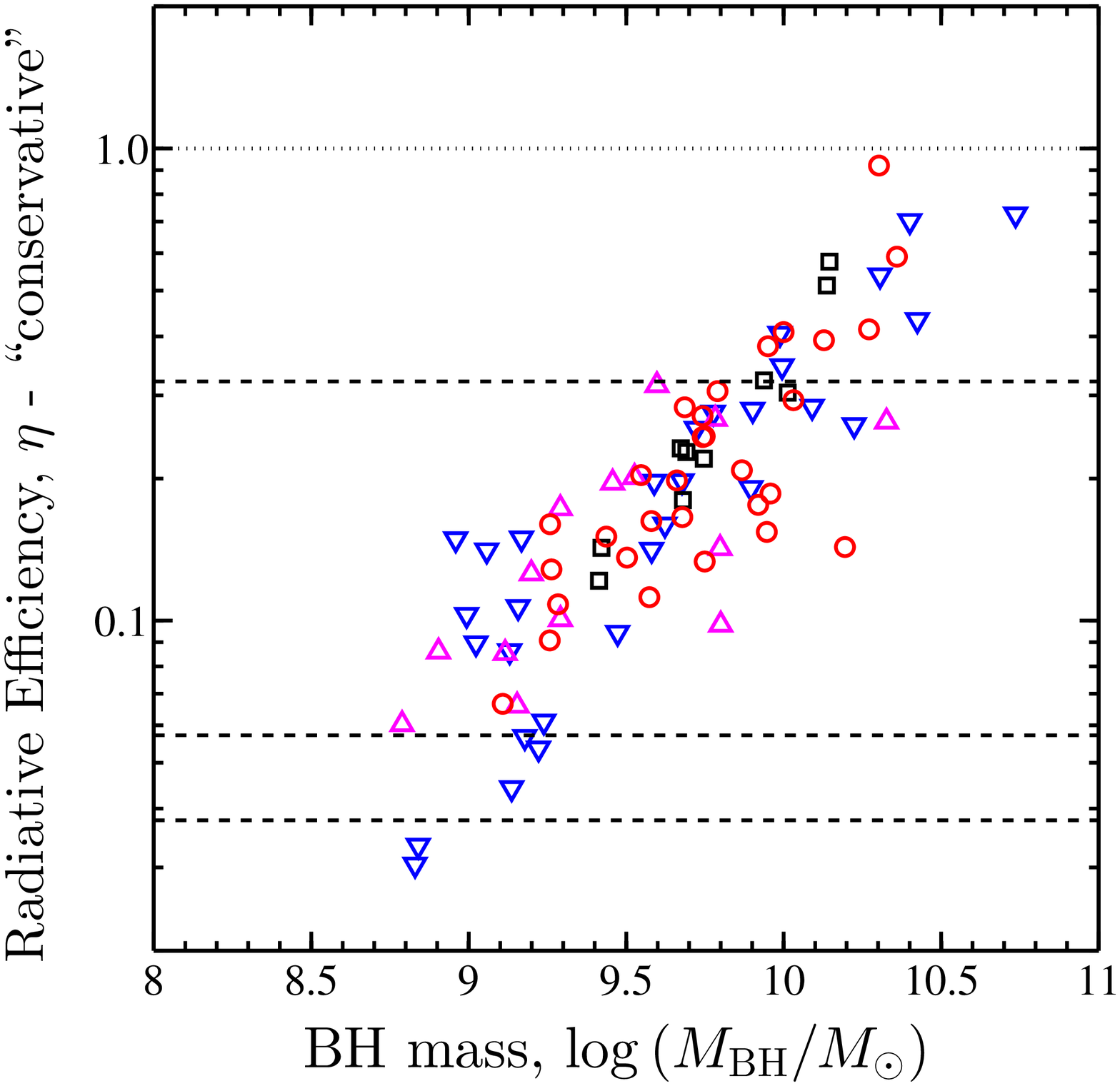} 
\hspace{0.3cm}
\includegraphics[width=0.48\textwidth]{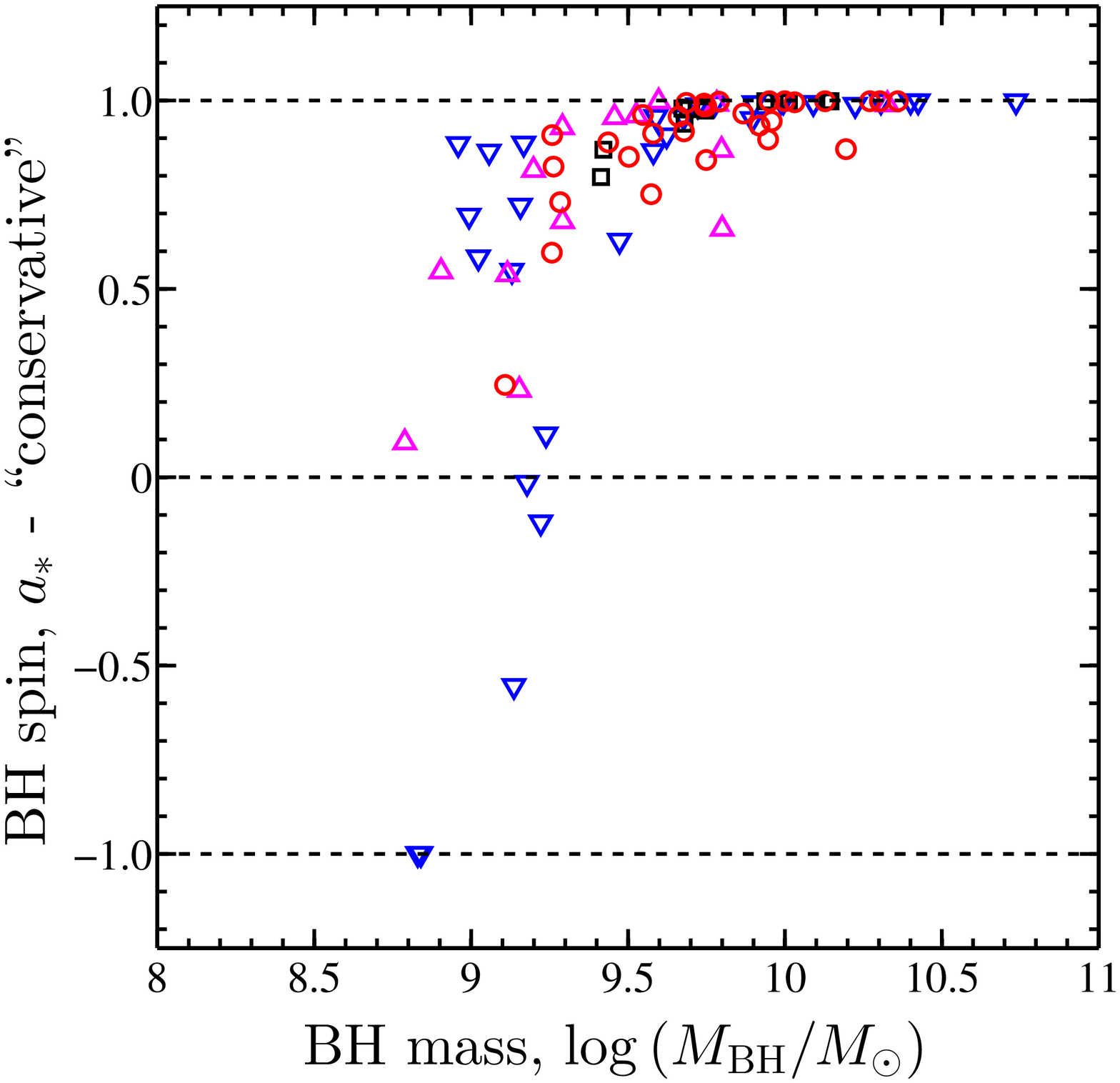} 
\caption{
\emph{``Conservative''} lower bounds on radiative efficiencies and BH spins, vs. \mbh, assuming a very small bolometric correction, $\Lbol=3\times\Lop$.
The panels descriptions are identical to Figure~\ref{fig:eta_a_vs_mbh_both_methods}.
}
\label{fig:eta_a_vs_mbh_low_Lbol}
\end{figure*}

\section{Results}
\label{sec:results}

Figure~\ref{fig:eta_a_vs_mbh_both_methods} presents the deduced radiative efficiencies and BH spins, derived from the ``standard'' estimates of \Lbol, and the \Lop-based estimates of \Maddot.
Clearly, the most massive BHs show very high radiative efficiencies and spins.
In fact, several sources exceed the well-known maximal radiative efficiency for a geometrically thin, optically thick AD (i.e., $\eta\gtrsim0.32$). 
These outliers, and the uncertainties associated with the estimation of \Lbol\ and \mbh,
motivate the use of more conservative estimates of \Maddot\ and \Lbol, following the alternative assumptions made in Section~\ref{sec:samples_analysis}. 

\begin{figure}[ht!]
\includegraphics[width=0.49\textwidth]{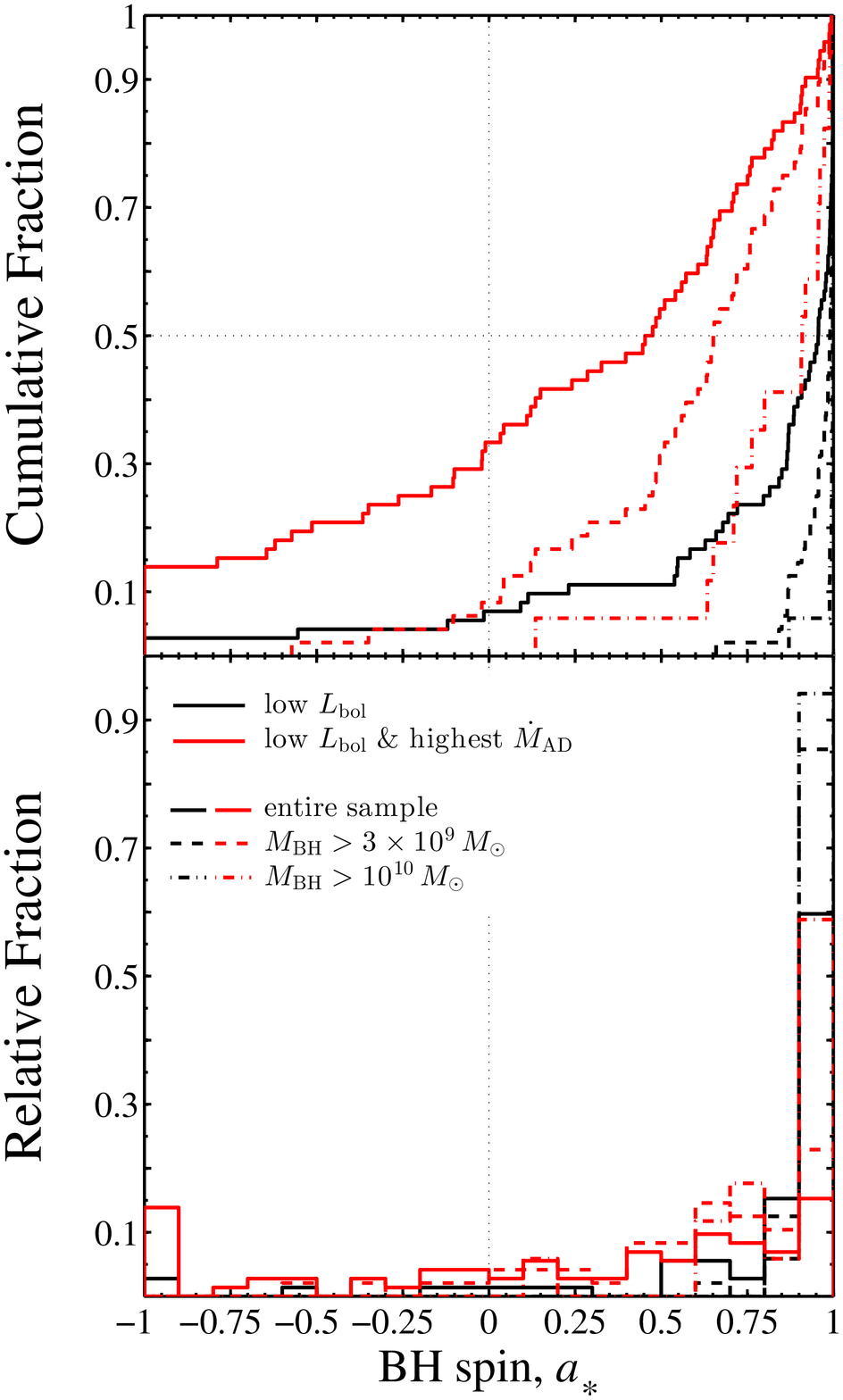} 
\caption{
Fractional (\emph{bottom}) and cumulative (\emph{top}) distributions of the BH spins for the high-redshift sources under study.
Black lines represent the ``conservative'' estimates, which assume $\Lbol=3\times\Lop$ and the \Maddot\ values derived through Equation~(\ref{eq:mdot_ad}) (using \Lop).
Red lines represent more extreme estimates, which further assume the \emph{highest} available estimate for \Maddot\ (including IR surveys and the DL11 prescription). 
In both cases, solid lines represent the entire sample, dashed lines represent the \NhiM\ sources with $\mbh>\MhiM\,\Msun$ and the dot-dashed lines represent the \NexM\ extreme sources with $\mbh>\MexM\,\Msun$.
Dotted vertical lines mark $\as=0$, and the horizontal line (top panel) marks the median value.
Note that even under the most extreme assumptions, the most massive BHs exhibit high spins, approaching $\as\simeq1$.
}
\label{fig:a_hist_cdf_cons_ext}
\end{figure}

As explained in Section~\ref{sec:samples_analysis}, the ``conservative'' calculations of \Maddot\ (and \re) assumed $\Lbol=3\times\Lop$, resulting in \re\ values which are systematically {\it lower} than the standard ones by a factor of $\sim$2.
Alternatively, this decrease in \Maddot\ could account for a situation where all the \mbh\ values were over-estimated, by a similar factor. 
Figure~\ref{fig:eta_a_vs_mbh_low_Lbol} presents the conservative estimates of \re\ and \as, versus \mbh. 
Clearly, even under the conservative assumptions, the basic qualitative result remains that many high-\mbh\ sources have high radiative efficiencies, and thus high spins. 
In particular, for both the \NhiM\ sources with $\mbh>\MhiM\,\Msun$ and the \NexM\ extreme sources with $\mbh>\MexM\,\Msun$, the conservative assumptions predict typical (median) spins of $\as\simeq0.98$. 
These median values and the overall distribution of \as\ derived under the conservative assumptions are illustrated in Figure~\ref{fig:a_hist_cdf_cons_ext}, 
which also presents the corresponding distribution for another set of yet more stringent assumptions. 
In this \emph{``extreme''} case, \re\ was estimated by using the conservative estimates of \Lbol\ (i.e., $\fbolopt=3$) \emph{and} the highest available value for \Maddot, using either Equation~(\ref{eq:mdot_ad}) with \Lop, the 2MASS- or \emph{WISE}-based luminosities, or the DL11 prescription.
As Figure~\ref{fig:a_hist_cdf_cons_ext} clearly shows, even under these extreme assumptions, the entire sample, and in particular the most massive BHs, exhibit high spins.
The \NhiM\ sources with $\mbh>\MhiM\,\Msun$ now typically have $\as\simeq0.65$, and the \NexM\ sources with $\mbh>\MexM\,\Msun$ have spins of $\as\simeq0.91$.
As noted in Section~\ref{sec:samples_analysis}, the uncertainties associated with estimating \re\ do not have a significant effect on these basic results. 
The apparent correlation between \re\ and \mbh\ in Figure~\ref{fig:eta_a_vs_mbh_low_Lbol} is mainly driven by the form of Equation~(\ref{eq:mdot_ad}) \cite[see also][]{Raimundo2012}. 
The fact that the derived \re\ for low-\mbh\ sources are conservative lower limits further weakens the case for any significant correlation.

As additional evidence for high BH spins, one may consider the opposite line of argument: assume a certain \as\ (or \re); combine it with the estimated \Maddot; and deduce the expected \Lbol\ (see also NT14).
For the sample in hand, assuming $\as=0.2$ (i.e., $\eta=0.065$) results in bolometric luminosities that are lower than the ``standard'' ones by more than 0.5 dex.
Thus, \emph{if} all the SMBHs in the sample had low spins, the typical value for \fbolopt\ should have been $\lesssim1.5$.
For the higher-\mbh\ sources, the typical values would imply $\fbolopt < 1$ (i.e., $\Lbol<\Lop$).
Such extreme values are in sharp contradiction with our current understanding of the UV-optical SED of AGNs \cite[e.g.,][and references therein]{Richards2006_SED,Runnoe2012}.
Thus, the scenario where all high-\mbh\ BHs have low spins can be rejected solely based on the estimated values for \Maddot.
Finally, as pointed out in \cite{LaorDavis2011_WLQs} and NT14, for extremely massive BHs with low spins the ADs would not produce enough ionizing radiation to produce the strong broad high ionization lines observed in their spectra (e.g., \CIV).

\section{Discussion and Conclusion}
\label{sec:discussion}

The analysis presented here provides indirect evidence that the most massive BHs at $z\sim1.5-3.5$ accrete with high radiative efficiencies, and therefore most probably have high spins. 
Even under conservative assumptions, the data in hand suggests that $\as \gtrsim 0.5$ for SMBHs with $\mbh\gtrsim\MhiM\,\Msun$ (Figure~\ref{fig:a_hist_cdf_cons_ext}).

As mentioned in Section~\ref{sec:intro}, several recent studies link the evolution of BH spin to the distribution of orientations of episodic accretion periods \cite[e.g.,][D13]{Volonteri2007_spin,King2008_spindown,Wang2009_rand_acc}.
These studies find that most SMBHs would ``spin-down'' to $\as\lesssim0.2$ if the orientations of the ADs are isotropic. 
Alternatively, the most massive BHs would ``spin-up'' to $\as\simeq1$ if the accretion is either coherent over very long timescales, or if the distribution of orientations is highly anisotropic (D13).
In this context, the present study provides strong evidence that the most massive active BHs at $z\sim1.5-3.5$ were ``spun-up'', 
in agreement with the predictions of relevant models (e.g., D13, \citealt{Volonteri2013_spin}).

The high masses of the SMBHs under study require prolonged accretion at high duty cycles \cite[e.g.,][]{Shemmer2004,Netzer2007_MBH,Netzer_Trakht2007,Dietrich2009_Hb_z2}. 
Indeed, the (exponential) growth times of most of the SMBHs studied here, $\tgrow$, estimated assuming $\left(\Mbhdot/\mbh\right)\propto \left(\left[1-\eta\right]/\eta\right) \lledd$ (and $\eta=0.1$), exceed the age of the universe (at the observed redshift, \tUni).
This tension may be somewhat resolved by invoking high-mass BH seeds, with $M_{\rm seed}\sim10^4-10^6\,\Msun$ \cite[see review by][]{Volonteri2010_rev}.
The high radiative efficiencies found in this work further challenge these scenarios, since the implied efficiency of \emph{mass assembly} onto the BHs is lower than what is commonly assumed, by factors of up to $\sim$1.5 (note that $\Mbhdot\propto\left(1-\eta\right)\Maddot$).  
A re-assessment of \tgrow\ assuming the \Maddot\ values found here
suggests growth times which are longer than the aforementioned ones by factors of $\sim1.5-6$. 
About 80\% of the SMBHs have $\tgrow>\tUni$, and about half have $\tgrow>4 \times \tUni$. 
Thus, even under the extreme assumption of $\mseed=10^6\,\Msun$, most of the SMBHs had to accrete continuously to obtain their high masses.
These growth times may be shorter if the most massive BHs were accreting at higher rates in their past, as suggested by several studies \cite[][]{Willott2010,DeRosa2011,Trakhtenbrot2011}, and/or if they had lower spins.

In conclusion, the long growth times inferred for the SMBHs under study support a scenario of prolonged mass accretion (over $\gtrsim$1 Gyr), which is most probably comprised of many shorter accretion episodes, with minimal gaps between them. 
These accretion episodes probably had isotropic (random) orientations, sustaining low spins and high mass growth efficiencies. 
Finally, a relatively prolonged accretion event (or several short, anisotropically-oriented events) could have spun the BHs up to $\as\simeq1$.
This is qualitatively consistent with a series of randomly oriented, yet anisotropic, accretion episodes, as suggested by several models (e.g., the ``$F=0$'' model of D13 and the ``coherent decline'' model of Volonteri et al. 2013).
Fundamentally different classes of SMBHs, such as those powering much fainter AGNs; sources that were not classified as broad emission line AGNs; or inactive SMBHs, may possibly have lower spins (see discussion in \citealt{Raimundo2012}, \citealt{LaorDavis2011_WLQs} and NT14).

Finally, since the SMBHs studied here are not expected to grow much further, the present results suggest that their descendants -- the most massive \emph{relic} BHs in the local universe \cite[e.g.,][]{McConnell2011_MBH_10} -- would also have high BH spins.
Some preliminary results, based on novel submillimeter very long baseline interferometry observations, support this picture: 
the SMBH in M87, with $\mbh=6.6\times10^9\,\Msun$ \cite[][]{Gebhardt2011_M87}, was shown to have $\as\gtrsim0.6$ \cite[][]{Doeleman2012_M87}.
Unfortunately, such observations can only probe very few objects.
The results presented here should be critically tested by observing the K$\alpha$ lines of the sources, 
which may only become feasible with the next-generation of X-ray telescopes \cite[e.g., ``\emph{Athena}+''; see][]{Dovciak2013}.

\bigskip

\acknowledgements
I thank the anonymous referee, whose feedback helped to improve the manuscript, and  H. Netzer, K. Schawinski and M. Koss for valuable comments and discussions.




\end{document}